Ferroelectric control of antiferromagnetism via coordination swapping in $A_2Mo_3O_8$ (A= Mn, Fe, Co)

Yaxin Gao,[1,2*] Sha Li[1] and Menghao Wu[2*]

[1]School of Physics and Mechanical Electrical & Engineering, Hubei University of Education, Wuhan, Hubei 430205, China.

[2]School of Physics, Huazhong University of Science and Technology, Wuhan, Hubei 430074, China.

*contact author: m201770145@hust.edu.cn (Y.G.); wmh1987@hust.edu.cn (M. W.)

Abstract  Transition metal molybdenum oxides $A_2Mo_3O_8$ (A= Mn, Fe, Co) are known to be polar magnets where A ions are located in either octahedrally or tetrahedrally coordinated sites. In this paper we predict that their polarizations can be reversed via swapping of two coordinations for A ions, giving rise to robust vertical ferroelectricity. Such unique ferroelectricity via coordination swapping can be used to control the Neel vector in altermagnetic $Fe_2Mo_3O_8$, while the large non-relativistic spin-splittings can be also altered by ferroelectric switching in Luttinger compensated magnetic $Mn_2Mo_3O_8$, both in the absence of net magnetization. However, their ultra-thin layers may possess net magnetizations that can be reversed upon ferroelectric switching. Our findings provide a new type of multiferroicity as well as a new avenue in control of antiferromagnetic spintronics.

Antiferromagnetic spintronics is appealing for applications in energy-saving and high-speed nonvolatile memory[1-6] since antiferromagnets are robust against external magnetic field with zero magnetization. Current-induced staggered torque may induce Neel vector switching for data writing[7], while more energy-saving approach via electrical field remains elusive. They generally lack spontaneous spin splittings in the band structures, making efficient data reading challenging via transport compared with ferromagnets that can be easily distinguished by tunneling magnetoresistance effect in tunneling junctions. Even non-relativistic spin splittings can be achieved in a few cases of so-called altermagnetism[8-13], the electrical control of such splittings remains to be a challenging task.

A possible approach is to utilize multiferroic couplings that makes the electrical control of magnetism possible. Such multiferroic couplings are long-sought in light of their high potential for multifunctional applications combining the merits of different ferroics. [14-19] Although high temperature ferroelectric-magnetic couplings allow for practical applications in the joined functionality of "magnetic reading + electric writing" for random-access nonvolatile memories, single-phase multiferroics are still rare due to the necessity of mutually exclusive physical features, such as the coexistence of insulating with empty $d$ shell for the conventional ferroelectric order and conducting with partially filled $d$ shell for magnetism. Multiferroic couplings remain elusive in prevalent multiferroics like $BiFeO_3$ with distinct origins of magnetism and ferroelectricity, while spin-driven ferroelectricity in so-called type II multiferroics are all weak with low Curie temperature.

In this paper we show predict antiferromagnetic spin splittings that can be controlled via ferroelectricity in transition metal molybdenum oxides $A_2Mo_3O_8$ (A= Mn, Fe, Co). In those hexagonal structures with space group $P6_3mc$, 3d magnetic A ions are located in both octahedrally and tetrahedrally coordinated sites while nonmagnetic $Mo^{4+}$ ions form trimers in $Mo_3O_8$ layers. They have garnered considerable interest due to their intriguing physical properties, including their prominent magnetoelectric effect[20], modulated magnetic ground states and optical diode effect[21], fluctuation-enhanced phonon magnetic moments[22] and magneton polarons[23]. It has been demonstrated that $Fe_2Mo_3O_8$ exhibit excellent diverse magnetically ordered ground states, which can be tuned by applying an external magnetic fields or Zn-doping[24-26]. Significant thermal Hall effect has also been reported, which was attributed to their lattice-spin interactions[27,28]. Here we show first-principles evidence of their large spontaneous polarizations that can be switched via coordination swapping of A ions driven by electric field, giving rise to robust ferroelectricity. Meanwhile the magnetism also depends on the coordination of A ions, rendering the desirable multiferroic coupling that remains elusive. The issues for data writing and reading in antiferromagnetic spintronics may also be resolved as they exhibit altermagnetism with switchable Neel vector, or Luttinger compensated magnetism with non-relativistic spin-splittings which can be also altered by ferroelectric switching.

Our theoretical calculations were performed using spin-polarized density functional theory (DFT) implemented in the Vienna *ab initio* Simulation Package (VASP 5.4) code[29,30]. The generalized gradient approximation (GGA) in the Perdew-Burke-Ernzerhof (PBE)[31] exchange-correlation functional and the projector augmented wave method[32] with a plane-wave basis were applied. Since standard DFT fails to describe the electronic structure for these strongly correlated transition metal oxides, the DFT+Hubbard U method implemented on basis of the Dudarev's propose was adopted, where the Hubbard repulsion $U_{eff}$ (U-J=5 eV) and $U_{eff}$ (U-J=4 eV) is imposed on Fe's 3d, Mn's 3d orbitals, respectively[33]. All the structures were relaxed using GGA+U approximations with the relaxation terminated once the total energies and residual atomic forces converged to < $10^{-6}$ eV and 0.01 eV/Å. The kinetic energy cutoff was set at 400 eV, and the Brillouin zone was sampled by the Monkhorst-Pack meshed method [34] with a 5×5×3 *k*-points grid at Γ center. The Berry phase method was employed to evaluate ferroelectric polarizations[35], and the ferroelectric switching pathways were obtained by using a generalized solid-state elastic band (G-SSNEB) method[36].

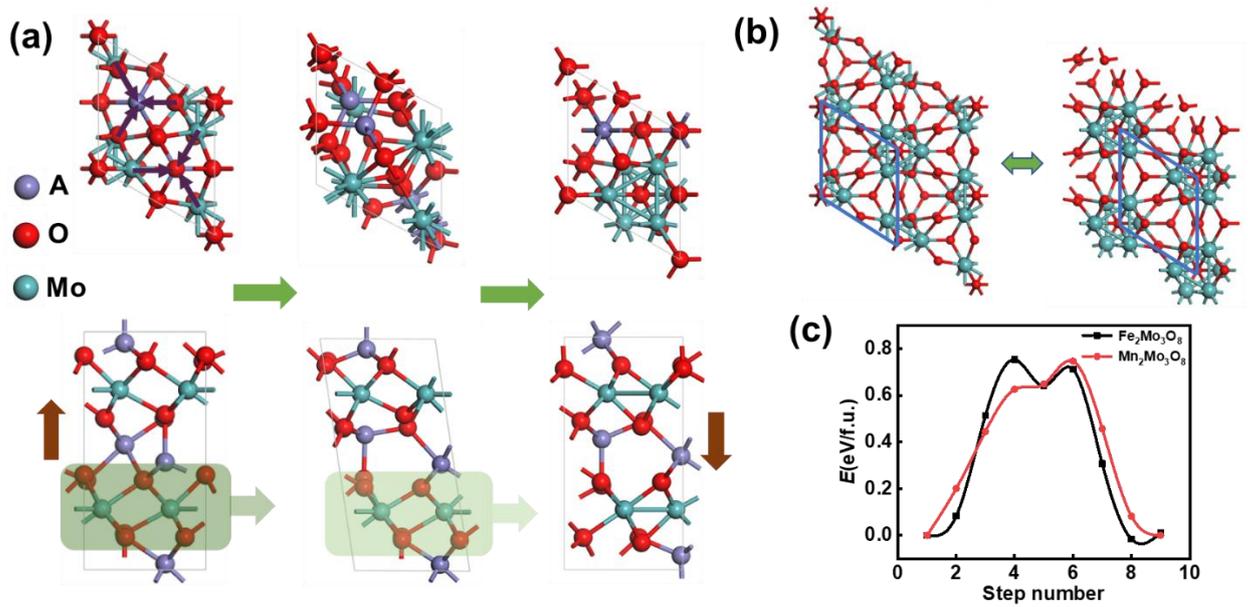

**Fig. 1** (a) The ferroelectric switching with the coordination configuration exchange of tetrahedral and octahedral $A^{2+}$ ion sites, where the polarization directions are denoted by brown arrows, and the translations of $Mo_3O_8$ layer are marked by green arrows. (b) The translation of the Mo trimers between initial states and final states. (c) Ferroelectric switching pathways of $Fe_2Mo_3O_8$ and $Mn_2Mo_3O_8$.

The geometric structures of $Fe_2Mo_3O_8$ are displayed in Fig. 1(a), where each primitive cell consists of two formula units and the optimized lattice constants (|a|:|b|:|c|=5.776:5.776:10.111) is consistent with experimental results (|a|:|b|:|c|=5.775:5.775:10.059).[37] For the two alternating nonequivalent sites in Fe cation layers intercalated between $Mo_3O_8$ layers, one is bonded to four oxygen atoms in a $FeO_4$ tetrahedron ($Fe_T$) and the other is coordinated by six oxygen atoms in a $FeO_6$ octahedron ($Fe_O$). The Fe layers are not strictly planar since the relative vertical displacements between $Fe_T$ sites and two adjacent $Mo_3O_8$ layers are different, and the symmetry breaking in crystal lattice gives rise to polarization of 21.2 μC/cm². Similar vertical ferroelectricity also emerges in other $A_2Mo_3O_8$, and the polarizations are respectively 18.9 and 23.1 μC/cm² for A= Mn and Co. As shown in Fig. 1(a), two identical polar states with different orientations for $A_T$ and opposite vertical polarizations, which are correlated by mirror symmetry *Mz*, can be interconvertible upon interlayer translation of $Mo_3O_8$ layers, accompanied by the swapping of coordination between $A_T$ and $A_O$ sites and reformation of trimerized $MoO_6$ octahedra (Fig. 1(b)). Considering the in-plane displacements of A ions, they may also exhibit in-plane ferroelectricity not complying with Neumann's principle similar to CuCrX$_2$(X=S, Se) with C$_{3v}$ symmetry,[16] which has been experimentally confirmed[38-40]. The pathway in Fig. 1(c) is computed by using the SSNEB method, revealing a moderate barrier of 0.37 eV per Fe ion for Fe$_2$Mo$_3$O$_8$, and 0. 38 eV per Mn ion for Mn$_2$Mo$_3$O$_8$.

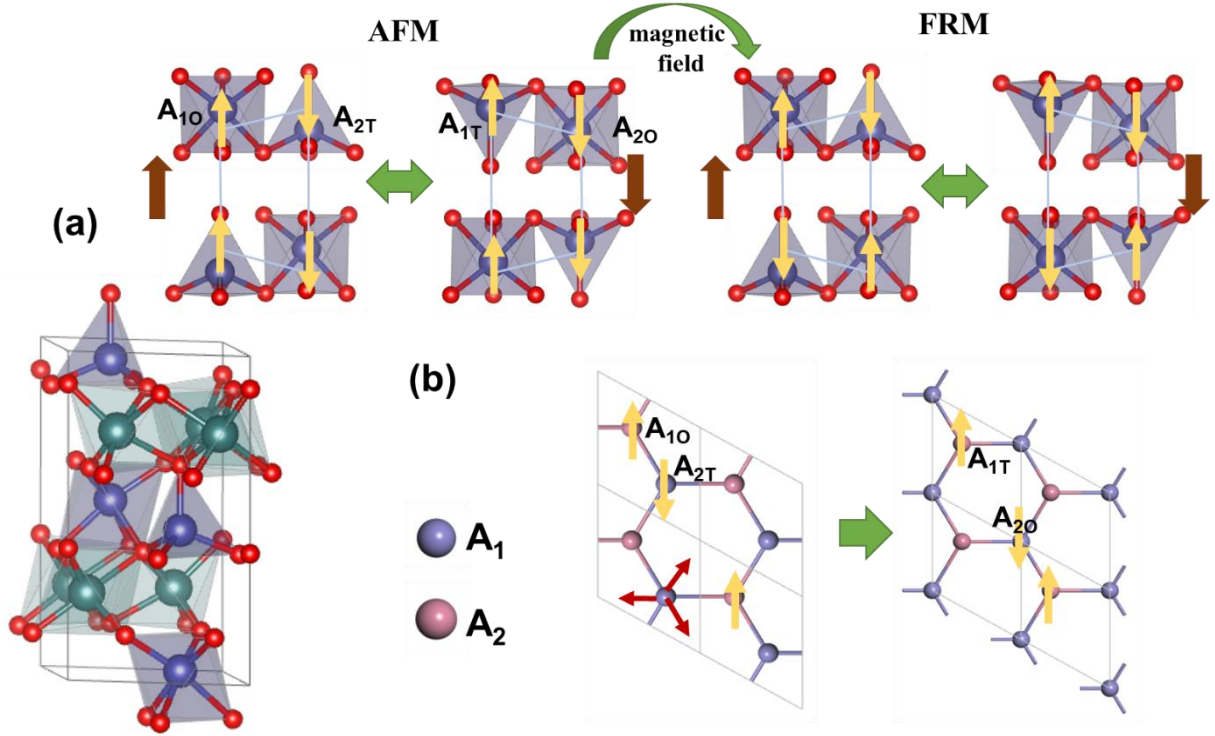

**Fig. 2** (a) Crystal structure of A$_2$Mo$_3$O$_8$ and schematic view of the AFM and FRM spin configurations upon ferroelectric switching, where yellow and brow arrows respectively denote the spin and polarization directions. (b) Overview of the spin configuration of each A layer upon ferroelectric switching.

Aside from ferromagnetic (FM) states, typical magnetic configurations for A$_2$Mo$_3$O$_8$ structure include antiferromagnetic (AFM, where all adjacent A$_T$ – A$_O$, A$_T$ - A$_T$ and A$_O$ -A$_O$ spin couplings are antiferromagnetic) and ferrimagnetic (FRM, where the spins of A$_T$ and A$_O$ sites are anti-parallel within each A layers, while the interlayer A$_T$ - A$_T$ and A$_O$ -A$_O$ spin couplings are ferromagnetic) state, which are displayed in Fig. 2 (a). Our DFT computation shows that the AFM state is the ground state for Fe$_2$Mo$_3$O$_8$, respectively 1 and 38 meV lower in energy compared with FM and FRM state, which accords with previous experimental reports, and the FRM state with very slightly higher energy can be tuned to the ground state by applying an external magnetic fields or Zn-doping[24-26]The magnetic moments of Fe$_T$ and Fe$_O$ in each Fe layer are respectively 3.704 and 3.740 $\mu_B$, giving rise to a net magnetization for each Fe layer antiferromagnetically coupled with its adjacent Fe layers. Similarly, AFM state is the ground state for $Co_2Mo_3O_8$, respectively 5 and 36 meV lower compared with FM and FRM state. In comparison, the FRM is the ground state for $Mn_2Mo_3O_8$, also consistent with the experimental results,[41] while AFM state is 7 meV higher in energy. However, the net magnetization of FRM $Mn_2Mo_3O_8$ is also exactly zero according to our calculations.

Distinct from most single-phase multiferrroics like $BiFeO_3$, herein the spin configurations of $A_2Mo_3O_8$ can be altered electrically via ferroelectric switching. Upon coordination swapping of $A_T$ and $A_O$, the magnetization of each A layer as well as the Neel vector will simultaneously switch, as shown by change of the honeycomb spin lattices of each A layer in Fig. 2(b). Meanwhile, non-relativistic spin splitting still emerge in those systems with zero net magnetization. For AFM state $Fe_2Mo_3O_8$, the two spin sublattices are connected by the combination of 180 degree rotation and translation transformation without inversion center between them, giving rise to altermagnetism[8] that can also be revealed by band structure in Figure 3(a). The band splitting is missing along the high-symmetry pathway in the plane $k_z$=0, while emerges along D (0.000, 0.000, 0.093) – U (-0.333, 0.667, 0.093) -P (0.000, 0.500, 0.093) -D parallel to Γ-K-M-Γ in the plane of $k_z$=0.3π/c planes[42].

The ground state of $Mn_2Mo_3O_8$ with zero net magnetization is actually Luttinger compensated magnetic phase[13], as the bandstructure in Fig. 3(b) turns to be a highly spin-polarized semiconductor. The number of occupied bands in spin-up and spin-down channel are exactly the same, giving rise to zero net magnetization even each $Mn_T$ and $Mn_O$ with different configurations are not likely to possess exactly the same magnetic moment. As analyzed by partial density of states (PDOS) in Fig. 3(b), the valance band maximum (VBM, in spin-up channel) and conduction band minimum (CBM, in spin-up channel) are distributed by $Mn_2$ in octahedrally coordinated sites initially. The large band splittings can be reversed upon ferroelectric switching with swapping of $Mn_T$ and $Mn_O$, and in the final state, the VBM and CBM respectively moves to spin-down and spin-up channel, which are distributed by $Mn_1$ transformed into octahedral coordination.

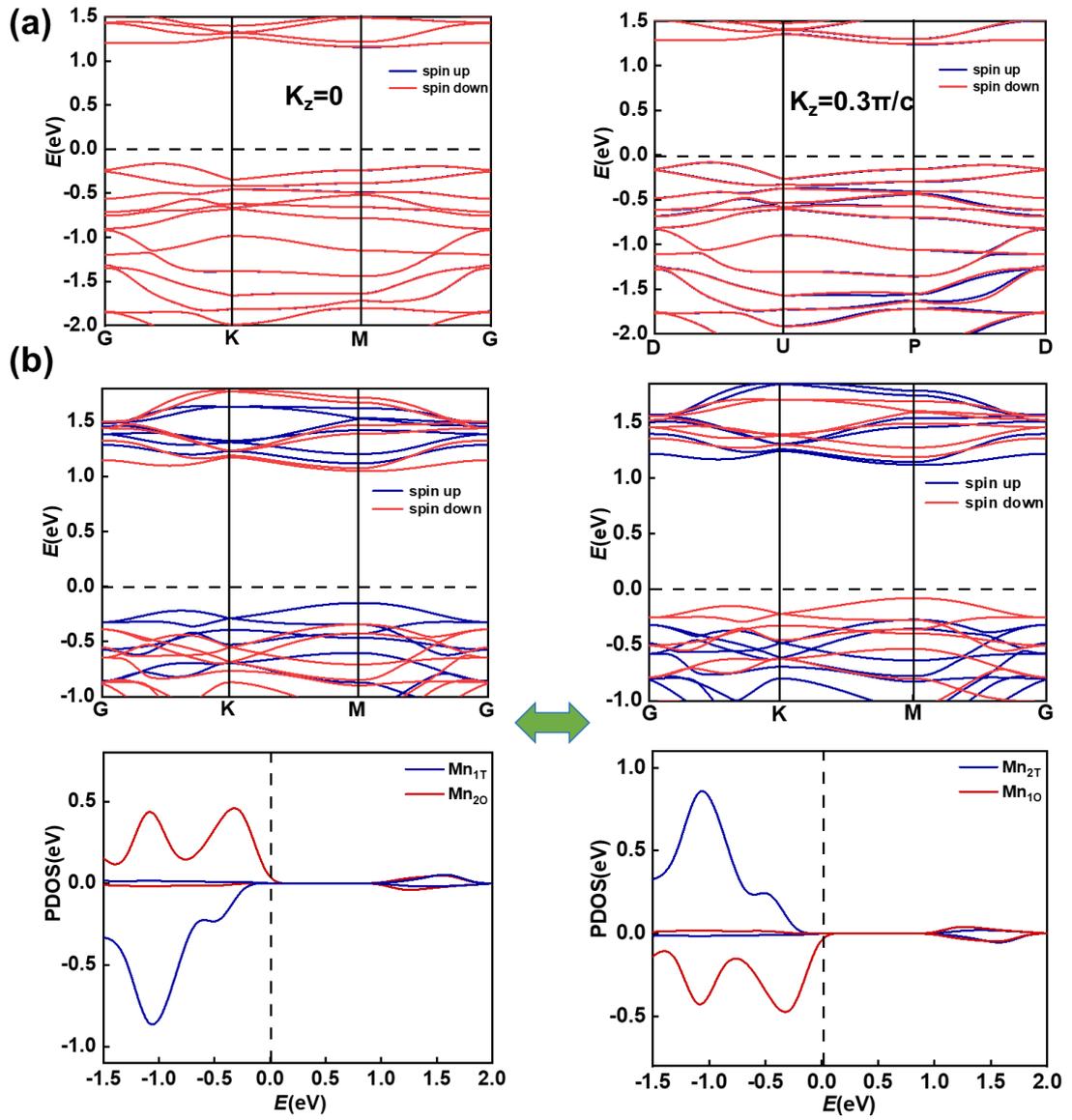

**Fig. 3** (a)The band structure of $Fe_2Mo_3O_8$ plotted on nodal ($k_z=0$) and off-nodal ($k_z=0.3\pi/c$) planes, where red and blue lines represent the spin-up and spin-down bands, respectively. (b) The change of band structure and PDOS of $Mn_2Mo_3O_8$ upon ferroelectric switching.

It is known that in ultrathin films of traditional ferroelectrics, the vertical polarizations disappear below critical film thickness due to the depolarizing filed. In comparison, the polarizations of $A_2Mo_3O_8$ structures contributed by the vertical displacements of $A_T$ ions can be hardly vanished. The thinnest structure of $Fe_2Mo_3O_8$ is shown at Fig. 4(a), which consists of two $MoO_6$ layers intercalated by $Fe_T$ and $Fe_O$ ions, denoted as $Fe_2Mo_6O_{16}$. In the magnetic ground state where the $Fe_T$ and $Fe_O$ sites are antiferromagnetically coupled, the difference in their magnetic moments leads to an uncompensated net magnetization of $0.25\mu_B$/f.u..

Although the $Fe_2Mo_6O_{16}$ system is metallic, the electrons are vertically confined and the vertical polarization may not be fully screened, similar to the ferroelectric metal $WTe_2$ bilayer reported previously.[43, 44] Its estimated vertical polarization of 2.42 pC/m is comparable to bilayer boron nitride (2.08 pC/m)[45], which is likely to be enhanced when surface terminations make the system more insulating and reduce the charge screening. The switching pathway shown in Fig. 4(b) reveals a reduced switching barrier of 0.32 eV per Fe ion. Upon the ferroelectric switching, the magnetic moment of $Fe_T$ and $Fe_O$ ions will be swapped, giving rise to a magnetic moment transfer between them as well as the switching of total magnetization $\sim 0.25\mu_B$/f.u., as illustrated by the change of spin distribution in Fig. 4(c).

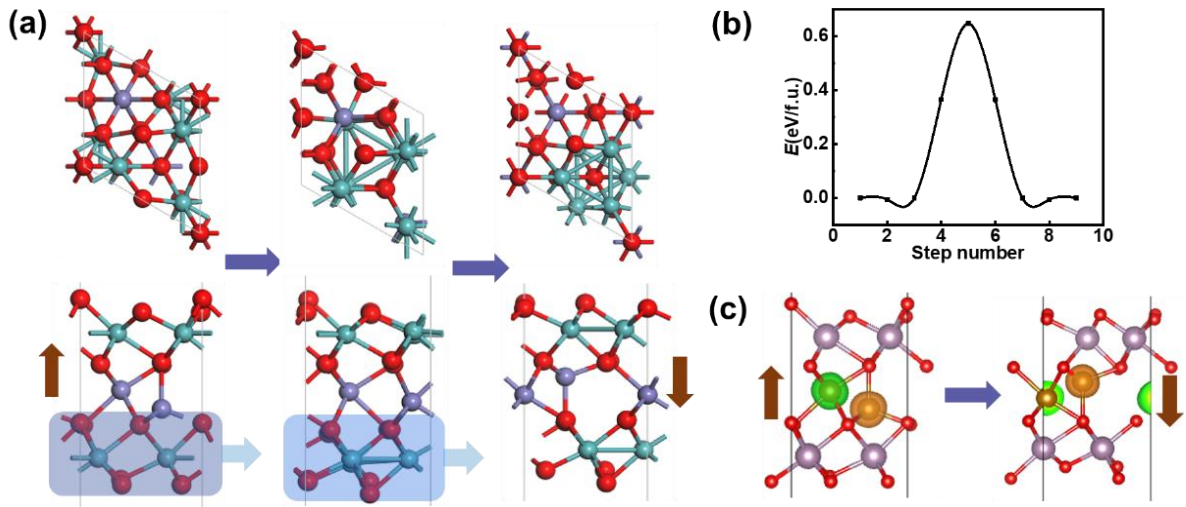

**Fig. 4** (a) Illustration of vertical ferroelectric switching for $Fe_2Mo_6O_{16}$ monolayer. (b) Computed ferroelectric switching pathway. (c) Change of spin distribution upon vertical ferroelectric switching, where green and orange respectively represent spin-up and spin-down density isosurfaces.

In summary, we predict a unique type of ferroelectricity in $A_2Mo_3O_8$ (A= Mn, Fe, Co) via coordination swapping of A ions. Even their net magnetizations are exactly zero, the Neel vector and non-relativistic spin-splittings can be altered upon ferroelectric switching via such coordination swapping. In their ultra-thin layers, their nonzero net magnetizations can be reversed upon ferroelectric switching. Such hitherto unreported multiferroicity in a single material may provide an efficient approach for data writing and reading in antiferromagnetic spintronics.

ACKOWNLEDGEMENT

This work is supported by National Natural Science Foundation of China (Nos. 22073034). We thank Prof. Jun-Ming Liu for helpful discussions.